\documentstyle[epsfig]{aipproc}

\begin{document}
\title{Long-Term {\it RXTE} Monitoring of the Anomalous X-ray Pulsar 1E~1048.1$-$5937}

\author{
Victoria M. Kaspi$^{* \dagger}$,
Fotis P. Gavriil$^*$,
Deepto Chakrabarty$^{\dagger}$,
Jessica R. Lackey$^{\dagger}$,
Michael P. Muno$^{\dagger}$
}

\address{$^*$Department of Physics, Rutherford Physics Building,
McGill University, 3600 University Street, Montreal, Quebec,
H3A 2T8, Canada\\
$^{\dagger}$Department of Physics and Center for Space Research,
Massachusetts Institute of Technology, Cambridge, MA 02139}

\maketitle

\begin{abstract}
We report on long-term monitoring of the anomalous X-ray pulsar
1E~1048.1$-$5937 using the {\it Rossi X-ray Timing Explorer}.
This pulsar's timing behavior is different from that of
other AXPs.  In particular, the pulsar shows significant
deviations from simple spin-down such that phase-coherent timing has
not been possible over time spans longer than a few months.  
We show that in spite of the rotational irregularities,
the pulsar exhibits neither pulse profile changes nor large pulsed flux
variations.  We discuss the implications of our results for AXP
models.  We suggest that 1E~1048.1$-$5937 may be a transition object
between the soft gamma-ray repeater and AXP populations, and the AXP
most likely to one day undergo an outburst.  
\end{abstract}

\section*{Introduction}

The nature of anomalous X-ray pulsars (AXPs) has been a mystery since
the discovery of the first example some 20 years ago.  Although it is
clear that AXPs are young neutron stars, it is not clear why they are
observable.  In particular, they show no evidence for possessing a
binary companion, making conventional accretion problematic.
Furthermore, given their spin periods and period derivatives, their
rate of loss of rotational kinetic energy is orders of magnitude too
small for these sources to be rotation-powered.  One important clue is
that two AXPs (and one AXP candidate) are clearly associated with
supernova remnants.  Although only five AXPs are known, their origin is
likely to be of great importance to our understanding of the fate of
massive stars and the basic properties of the young neutron star
population.  For an excellent recent review of these objects, see
\cite{mer99}.

Currently there are two models to explain AXPs.
One model proposes that AXPs are young, isolated, highly
magnetized neutron stars or ``magnetars'' \cite{dt92a,td96a}.
High magnetic fields ($10^{14}-10^{15}$G) are inferred from their 
spin-down under the assumption of magnetic dipole braking, 
as well as by association with the soft gamma repeaters
(SGRs) which show AXP-like pulsations in quiescence \cite{kds+98,ksh+99}, 
and are thought
to have high magnetic fields for independent reasons \cite{td96a}. 
The second model of AXP emission is that they are powered by
accretion from a fall-back disk of material remaining from the
supernova explosion \cite{chn00}.

One way to distinguish between these classes of models may be through
timing observations.  In the magnetar model, relatively smooth
spin-down should be expected, punctuated by occasional abrupt spin-up
or spin-down events or ``glitches,'' as well as low-level,
long-time-scale deviations from simple spin-down, or ``timing noise.''
Both phenomena are well known among young radio pulsars (e.g.
\cite{lyn96}), although their physical origins in magnetars may be
different given the much larger inferred magnetic field.
However, according to the magnetar model, no extended spin-up should 
be seen.  On the other hand, accretion power is usually
associated with much noisier timing behavior, which can be correlated
with spectral, luminosity, and pulse morphology changes.  In addition,
some accreting binary systems undergo extended ($\sim$years) episodes
of spin-up, although these generally seem to alternate with long
intervals of spin-down \cite{bcc+97}.

1E~1048.1$-$5937 is a 6.4~s AXP in the Carina region \cite{scs86}.
It exhibits no evidence for any binary companion, as no Doppler
shifts of the pulse period are seen \cite{mis98}, and no optical
counterpart to a limiting magnitude of $m_V \sim 20$ has been detected
\cite{mcb92}.  The pulsar's spectrum, like those of other AXPs, is
well described with a two component model consisting of a soft black
body with a power-law tail \cite{opmi98}.  Occasional monitoring 
observations over more than 20
years show that the pulsar is spinning down, though significant
deviations from a simple spin-down model have been noted
\cite{opmi98,pkdn00,bss+00}. The paucity of data thus far
has made it impossible to unambiguously identify the origin of the
deviations.

Here we report on our monthly {\it Rossi X-Ray Timing Explorer} ({\it RXTE})
monitoring of 1E~1048.1$-$5937 in which we have attempted long-term phase-coherent 
timing like that achieved for other AXPs  \cite{kcs99}. 
The results described here are reported in more detail in \cite{kgc+01}.

\section*{Observations and Results}

The observations we report on were made with {\it RXTE}'s Proportional
Counter Array (PCA; \cite{jsg+96}). Observations of 3--6 ks in length of 
1E~1048.1--5937 were made on a monthly basis during 1996 November
-- 1997 December and 1999 January -- 2000 August.
In addition, we used archival observations from 
1996; these generally had longer integration times than the other data
sets. To minimize use of telescope time, our monitoring data consist of
brief (usually 3~ks) snapshots of the pulsar.  These snapshots suffice
to measure pulse arrival times for a phase-coherent timing analysis
to good precision.  However, for any one epoch, the measured period has
typical uncertainty $\sim 3$~ms, quite large by normal timing
standards.  Thus, our snapshot method of measuring pulse arrival times can determine
spin parameters with extremely high precision only when phase
coherence can be maintained.  For details regarding this timing
procedure, see \cite{kgc+01}.
The snapshot observations are always, however, useful for monitoring
the source pulsed flux and pulse morphology (see below).

\subsection*{Timing}

We maintained unambiguous phase coherence for 1E~1048.1$-$5937 in our
monthly observations from 1999 January 23 through 1999 November 15.  We
required a fourth-order polynomial to characterize the 17 pulse arrival
times obtained in this span.  
These results alone clearly imply that the rotational behavior of
1E~1048.1$-$5937 is quite different from that of AXPs
1E~2259+586 and RXS~J170849.0$-$400910.  Those AXPs exhibit much
more stable rotation on comparable and even longer time scales,
that is, terms of higher order than $\dot{\nu}$ are very small or negligible for
those pulsars on time scales of over a year \cite{kcs99}. 
The span 1999 January through November
represents the longest over which we can phase-connect timing data from 
1E~1048.1$-$5937.  Investigating 
archival {\it RXTE} data going back to 1997 for 1E~1048.1$-$5937, 
we find timing results that are similar those obtained in our recent
monitoring program, namely we are able to maintain phase coherence
only over few-month intervals.

We can compare our pulse ephemerides with measurements of pulse
frequency made over the past 20~yr in order to look for long-term
trends.  Figure~1 shows the spin history of
1E~1048.1$-$5937 with previously measured spin frequencies plotted as points with
their corresponding 1$\sigma$ error bars.  Data were taken from a variety
of sources \cite{opmi98,pkdn00,bss+00}.  Our {\it RXTE} timing results are plotted
as lines representing separate, short, phase-connected segments. 
The dotted line represents an
extrapolation of the $\nu$ and $\dot{\nu}$ from the 1999 coherent fit.
The lower plot shows the same data set with the linear term subtracted
off.  This ephasizes deviations from the simple linear trend.

\begin{figure}[t!] 
\centerline{\epsfig{file=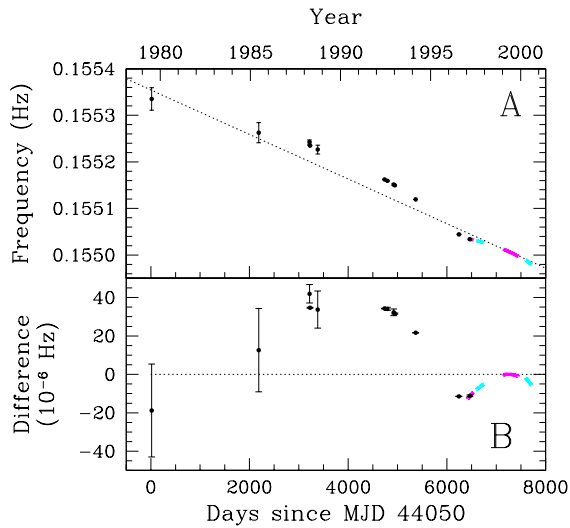,height=3in,width=3in}}
\vspace{10pt}
\caption{Spin history for 1E~1048.1$-$5937.  The points represent
past measurements of the frequency of the pulsar. The solid lines
represent the {\it RXTE} phase-connected intervals.  See \protect\cite{kgc+01} for
details. Panel A shows the observed frequencies over time.  The dotted
line is the extrapolation of the $\nu$ and $\dot{\nu}$ of the 1999
phase-coherent ephemeris.  Panel B shows the
difference between the ephemeris indicated by the dotted line and the data points.}
\end{figure}

\subsection*{Pulsed Flux and Pulse Morphology}

In accreting systems in which the neutron star is undergoing spin-up,
changes in torque should be correlated with changes in X-ray flux.
AXPs are spinning down, however.
Chatterjee et al. (2000) suggest that AXPs might be spinning down in
the propeller regime due to accretion from a fall-back disk.
In that case, although the physics of the propeller regime is not
well understood, it is still likely that $L_x$ should be correlated
with torque \cite{kgc+01}.

Given the large field-of-view of the PCA and that the bright, nearby, but
unrelated source $\eta$ Carinae exhibited large flux changes over the
course of our observations, direct flux measurements of
1E~1048.1$-$5937 could not be made with our {\it RXTE} data.  Instead,
we determined the pulsed component of the flux by using off-pulse
emission as a background estimator.  This renders our analysis
insensitive to changes in the fluxes of other sources in the
field-of-view.

The results are shown in Figure 2.
We find no large pulsed flux variations.
The $\chi^2$ strictly speaking does suggest some low-level variability;
longer individual observations are clearly necessary to verify
this is the case.  However, as we discuss below,
the pulsed flux is certainly much more
stable than previous analyses have suggested \cite{opmi98}.

\begin{figure}[t!] 
\centerline{\epsfig{file=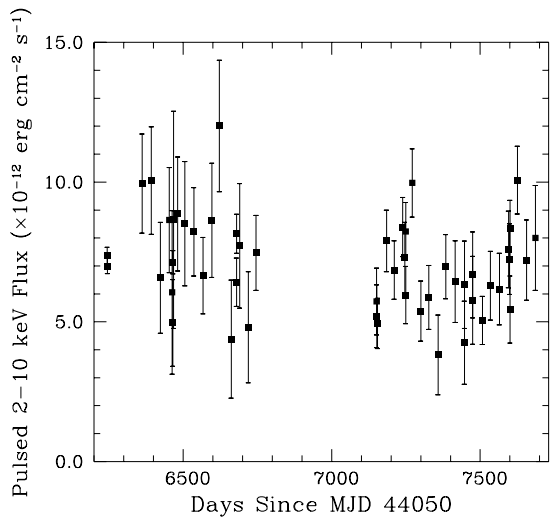,height=3in,width=3in}}
\vspace{10pt}
\caption{Pulsed flux time series in the 2--10~keV band
for {\it RXTE} observations of 1E~1048.1$-$5937.}
\end{figure}

We have also used the {\it RXTE} data to search for pulse profile changes,
as many accretion-powered pulsars exhibit significant
changes in their average pulse profiles.  Such changes can be
correlated with the accretion state, and hence accretion torque and
timing behavior \cite{bcc+97}.  Furthermore, X-ray pulse profiles from
the SGRs 1806$-$20 and 1900+14 have shown differences at different
epochs depending on time since outburst \cite{kds+98,ksh+99}.
However, we find no significant changes in the pulse profile morphology in any
of the {\it RXTE} observations of 1E~1048.1$-$5937.

\section*{Discussion and Conclusions}

Long-term {\it RXTE} monitoring of the AXP 1E~1048.1$-$5937 has shown
it to be a much less stable rotator than other AXPs, yet its pulse
profile and pulsed flux are stable.  
Previously, Oosterbroek et al. (1998) compiled flux data from a
variety of different X-ray instruments that observed 1E~1048.1$-$5937.  That
compilation suggested that the pulsar shows variability by over a
factor of $\sim$5 on time scales of a few years.  The reality of those flux changes
is not supported by our results.  
One caveat is
that we measure pulsed flux, while they report flux, so the results
could be reconciled if the pulsed fraction is variable. 

In the context of the magnetar model, we note that the timing behavior of
1E~1048.1$-$5937
is somewhat similar to that observed for the soft gamma repeaters SGR 1806$-$20
and 1900+14 \cite{mrl99,wkf+00,wkp+00}. 
However, as the stable flux time series (Fig. 2)
for the AXP shows, it has not undergone any outbursts.  This can perhaps be
understood in terms of persistent seismic activity and small-scale crustal
fractures \cite{tb98} or low amplitude toroidal modes resulting in angular momentum loss 
following crustal twisting fractures \cite{dun00}.

1E~1048.1$-$5937 is unusual among AXPs for reasons other than just its
timing behavior.  In particular, 
it shows the highest ratio of blackbody to total flux (once
energy band is accounted for), and the largest pulsed fraction.  In
addition, it has the lowest photon index for the power-law tail in
its spectrum of any AXP, which makes it the closest to those
measured in the X-ray band for SGRs 1806$-$20 and 1900+14.  Further,
the thermal component of 1E~1048.1$-$5937's spectrum has the highest
temperature (0.64~keV) of any AXP.  This temperature is  
comparable to that seen for SGR 1900+14
post-burst, 0.62~keV \cite{wkp+00}.  It therefore could be the case
that 1E~1048.1$-$5937 is a transition object between the populations of
AXPs and SGRs, and the AXP most likely to one day undergo an outburst.

In the context of accretion models, perhaps the best source with which
to compare 1E~1048.1$-$5937 is 4U~1626--67, a 7.7~s accreting pulsar in
a 42-min binary with a low-mass companion.  Although 1E~1048.1$-$5937
is noisy by AXP timing standards, its noise is comparable in strength
to that of 4U~1626$-$67 \cite{cbg+97}.  Still, we regard the case for
1E~1048.1$-$5937 as an accreting binary, even with a very low-mass
companion, as weak, given the other evidence against this hypothesis,
namely, its much softer spectrum than other accreting binaries, the
absence of pulsed flux or pulse morphology changes correlated with the
timing behavior, and the spin-down over some 20~yr.  It is more
difficult to dismiss the possibility that 1E~1048.1$-$5937 is accreting
from a supernova fall-back disk, since there is not yet a consensus on
the properties such a disk would have or on the expected timing and
variability properties of the pulsar.  However, one expectation is that
such a disk would be a significant emitter in the optical and infrared,
Future optical/IR observations
following a more precise localization using the {\it Chandra X-ray
Observatory} could test the fallback disk model.

\end{document}